\documentstyle[aps,preprint,prb,psfig]{revtex}

\begin{document}
\draft
\title{Microscopic study of electrical transport through individual 
molecules with metallic contacts: II. Effect of the interface structure}

\author{ Yongqiang Xue ~\cite{Xue} and Mark Ratner}
\address{Department of Chemistry and Materials Research Center, 
Northwestern University, Evanston, IL 60208}
\date{\today}
\maketitle

\begin{abstract}
We investigate the effect on molecular transport due to the different 
structural aspects of metal-molecule interfaces. The example system chosen 
is the prototypical molecular device formed by sandwiching 
the phenyl dithiolate molecule (PDT) between two gold electrodes with 
different metal-molecule distance, atomic structure at the metallic 
surface, molecular adsorption geometry and with an additional hydrogen end 
atom. We find the dependence of the conductance on the metal-molecule 
interface structure is determined by the competition between the 
modified metal-molecule coupling and the corresponding modified energy 
level lineup at the molecular junction. Due to the close proximity of the 
HOMO of the isolated PDT molecule to the gold Fermi-level, this leads to 
the counterintuitive increase of conductance with 
increasing top metal-molecule distance that decreases only after the 
energy level line up saturates to that of the molecule chemisorbed on the 
substrate. We find that the effect on molecular transport from adding an apex 
atom onto the surface of a semi-infinite electrode is similar to 
that from increasing the metal-molecule distance. The similarity is 
reflected in both the charge and potential response of the molecular 
junction and consequently also in the nonlinear transport characteristics. 
Changing the molecular adsorption geometry from a three-fold to an atop 
configuration leads to slightly favorable energy level lineup for the 
molecular junction at equilibrium and consequently larger conductance, 
but the overall transport characteristics remain qualitatively the same. 
The presence of an additional hydrogen end atom at the top metal-molecule 
contact substantially affects the electronic processes in the molecular 
junction due to the different nature of the molecular orbitals involved and 
the asymmetric device structure, which leads to reduced conductance and 
current. The results of the detailed microscopic calculation can 
all be understood qualitatively from the equilibrium energy level lineup 
and the knowledge of the voltage drop across the molecular junction at 
finite bias voltages.     
\end{abstract}

\date{\today}

\section{Introduction\label{S1}}
Transport measurement of molecular junctions typically involves binding of 
molecules onto metallic electrodes through an appropriate end group. 
Often the contact with one of the electrodes is established by 
self-assembling the molecules on a single-crystal substrate. The other 
contact can be formed by vacuum deposition of a top metal layer, by using 
a scanning tunneling microscope (STM) tip or a conducting atomic force 
microscope 
tip~\cite{JGA00,Metzger97,Reed992,Datta97,Xue991,Hong00,Bao02,Lindsay01}. 
Contacts to the electrodes can also be made by using atomic-scale break 
junctions~\cite{Reed97,Mccuen02,Park02,Weber02,Shash02}. 
The metal-molecule interface can therefore differ in metal-molecule 
distance, adsorption geometry and atomic structure of the metallic 
electrodes, or chemically in the types of the 
end groups and metals used. In most experiments, 
the structure of the metal-molecule interface is not known and cannot 
be controlled easily. This has hindered identifying the correct conduction 
mechanism through the molecular junction, since it is not clear whether 
the measured transport characteristics are intrinsic to the molecules or 
are due to features of the metal-moleule interface that might be 
non-reproducible from sample to sample~\cite{Weber02}. 
The purpose of this second paper on microscopic study of 
single-molecule electronics is to elucidate the effect on molecular
transport due to the different structural aspects of the 
metal-molecule interface. In the next paper, we discuss the effect on 
molecular transport due to the chemical aspects of the metal-molecule 
interface.  

Since for a given molecule-metal combination, the atomic-scale structure 
of the metal-molecule interface differs in numerous ways, it is 
not useful to discuss exhaustively such differences and their effect on 
the transport characteristics without reference to specific transport 
measurements. Instead this work aims at identifying the key conceptual 
issues involved and demonstrates the use of such concepts through 
detailed microscopic study of selected aspects of the metal-molecule 
interface. Such concepts should then be useful in giving a clear indication 
of whether given device characteristics originate from features of 
the metal-molecule interface structure. In the first paper of this 
series~\cite{Xue1} (refereed to as {\bf I} in this work), we have 
identified two key factors for 
understanding the transport characteristics of a molecular junction: 
the equilibrium energy-level lineup and the nonequilibrium charge/potential 
response to the applied bias. Electronic 
processes at the metal-molecule interface play different roles in 
determining these two factors: (1) At equilibrium, the symmetry and the 
magnitude of the metal-molecule orbital overlap determine the 
capabilities of the molecular states to function as effective conduction 
channels, while the coupling-induced charge and potential perturbation 
determine the shift of the molecular level relative to the metal 
Fermi-level; (2) Out of 
equilibrium, the asymmetry of the coupling at the source-molecule and 
drain-molecule interfaces determines the net charge flow into the molecule, 
but the spatial distributions of the charge response and the voltage drop 
are determined by the potential landscape across the entire 
molecular junction. At not too high bias voltage, this affects mainly the 
shift of the molecular levels with bias voltage (see {\bf I}). Our 
discussion will therefore focus on analyzing how the 
different aspects of the interface structure affect the above two types 
of electronic processes. 

Since the effect of the metal-molecule interface structure on the device 
characteristics is seen most clearly for symmetric molecules, we will use 
as example the prototypical molecular device formed by 
attaching the phenyl dithiol molecule (PDT) onto two gold electrodes 
through the end sulfur atoms. Since in many experiments on molecular 
transport, the molecules are self-assembled on a single-crystal 
substrate and the structure of the molecule-substrate contact can be 
considered well-defined, we start from the reference interfacial 
configuration where the molecules form symmetric contact with two 
semi-infinite gold $<$111$>$ electrodes (see {\bf I}) and investigate 
the change in its transport characteristics due to structural differences 
in terms of the metal-molecule distance, atomic structures of the metallic 
surfaces and the adsorption geometry. Since in practice it is not clear 
whether the end hydrogen atoms are desorbed upon electrode 
contact~\cite{Xia,Tour98}, we will 
also investigate the effect on current transport due to an additional end 
hydrogen atom at the top metal-molecule interface.      
The theoretical approach we use in elucidating the effect of 
the metal-molecule interface structure were discussed in detail in {\bf I} 
and also elsewhere~\cite{Xue01,Xue021}. We use the same modeling methodology 
as described in {\bf I}. In particular, we have used the 
BPW91-parameterization of the spin-density-functional 
theory~\cite{Becke88GGA,PW91,JG89} and the \emph{ab initio} 
pseudopotential~\cite{KBT} with the corresponding 
energy-optimized gaussian basis sets~\cite{Stevens84,Note1}. The 
calculation is performed using a modified version of GAUSSIAN98~\cite{G98}. 
We will focus on the results of the computation and the conceptual 
understanding derived from them (we use atomic units throughout the paper 
unless otherwise noted).  

\section{The effect of the metal-molecule distance \label{S2}}
In the reference device structure, the molecule forms symmetric contact 
with two semi-infinite gold $<$111$>$ electrodes, and the molecule 
sits on top of the center of the triangular gold pad (see {\bf I}). 
The end sulfur atom-metal surface distance at both interfaces 
is $1.9( \AA)$. We first 
consider the effect on current transport from increasing the 
metal-molecule distance for the right contact (denoted top contact 
from here on. The left contact is denoted the substrate contact). 
Increasing the top metal-molecule distance reduces the strength of orbital 
coupling across the interface without changing its symmetry, which also 
affects the equilibrium energy-level lineup since 
it changes the magnitude of the charge and potential perturbation across 
the molecular junction. The conductance of the metal-molecule-metal 
junction will depend on the competition between these two factors. This is 
demonstrated by examining both the equilibrium transmission versus energy (TE) 
characteristics and the projected molecular density of states (PDOS). The 
calculated TE characteristics, the PDOS and the current/conductance-voltage 
characteristics of the molecular junction for symmetric contact case 
and for cases where the top metal-molecule distance increased 
by $\Delta L=0.5,1.0,1.5,2.0( \AA)$ are shown in 
Figs. (\ref{xueFig2-1-1}-\ref{xueFig2-2}). To check that the calculation is 
not sensitive to small changes in the metal-molecule distance, we have 
also calculated the transport characteristics of the molecular junction for 
$\Delta L=0.1,0.2 (\AA)$, which show smooth deviation from the reference 
symmetric contact case. 

Increasing the top metal-molecule distance up to $1( \AA)$ 
\emph{increases} the conductance of the molecular junction. Although this 
result is counterintuitive, it is readily understood by examining the 
energy-level lineup at the metal-molecule-metal junction. The calculated 
HOMO level of the isolated PDT molecule in the spin-singlet state 
(with the end hydrogen atoms removed) is $-5.33(eV)$, very close to the 
metal Fermi-level $-5.31(eV)$. For the symmetric contact case, the charge 
transfer and the associated electrostatic potential change across the 
interface push it down to about $1(eV)$ below the Fermi-level (deduced 
from the corresponding peak positions in the PDOS plot, see {\bf I}). 
Increasing the top metal-molecule 
distance will therefore move the HOMO up to closer alignment with the metal 
Fermi-level, which overcompensates the decreased coupling strength in 
determining the transmission coefficient at the Fermi-level and 
correspondingly the low-bias  conductance. This trend continues as we 
increase $\Delta L$ from $0$ to $1.0( \AA)$. Note that although the 
transmission coefficient at the Fermi-level increases, the transmission 
through the middle of the HOMO-LUMO gap decreases monotonically with  
increase of the top metal-molecule distance as expected. We can also 
see that the effect on the alignment of the LUMO level is much weaker than 
that of the HOMO as we increase $\Delta L$ from $0.5( \AA)$ to $1.0( \AA)$, 
since the electron distribution of the LUMO is localized in the interior of 
the molecule (see {\bf I}).      
As we further increase the top metal-molecule distance to 
$\Delta L=1.5 (\AA)$, the coupling across the top metal-molecule interface 
becomes so weak such that its effect on the energy level lineup 
saturates, i.e., the energy level lineup approaches that of the molecule 
chemisorbed on the substrate. Further increasing the top 
metal-molecule distance will then reduce the transmission coefficient 
and the low-bias conductance of the molecular junction. Going from 
$\Delta L=1.5( \AA)$ to $\Delta L=2.0( \AA)$, the peak positions 
in both TE and PDOS plots corresponding to the HOMO and LUMO level do not 
change much, but the transmission coefficient is reduced across the entire 
energy spectrum (Fig. (\ref{xueFig2-1-2})). 

As the top metal-molecule distance increases, the equilibrium transmission 
characteristics also change substantially. In particular, the 
double-peak structure corresponding to resonant transmission through the 
LUMO and LUMO+1 is reduced to a single peak for transmission through the 
LUMO+1 state. The transmission probability away from the two 
peaks at HOMO and LUMO+1 is reduced rapidly with the increasing top 
metal-molecule distance. Examinations of the corresponding LDOS show that 
the LUMO is mainly localized on the peripheral hydrogen atoms, which leads to 
negligible orbital overlap with the top metal states and rapid reduction 
of the transmission probablilty with increasing top metal-molecule distance. 
The HOMO and LUMO+1 levels instead have large weights on both end sulfur 
atoms, so their transmission probability decreases much slower with the 
increasing top metal-molecule distance. Since transport is dominated by  
by tunneling through HOMO in the bias voltages studied here, the changes 
in the transmission through the unoccupied molecular states do not affect 
the device characteristics.    
 
As $\Delta L$ increases, the asymmetry of the I-V and G-V characteristics 
with respect to bias polarity also increases (Fig. (\ref{xueFig2-2})), 
since the bias-induced modification of molecular states (the Stark effect) 
differes at different bias polarities due to the asymmetry (the bias 
polarity is chosen such that positive bias voltage 
corresponds to electron injection from the top contact). 
This is clear from the bias-dependence of the 
molecular levels in Fig. (\ref{xueFig2-3}). At large top metal-molecule 
distance, the molecule couples strongly to the substrate metal but 
very weakly to the top metal. The calculated I-V and G-V 
characteristics (Fig. (\ref{xueFig2-2})) correspond to that obtained for   
tunneling through a chemisorbed molecule, e.g., measured using a STM tip 
with a large vacuum gap~\cite{Ho98}. At first sight, this suggests that 
the shift of the molecular levels with applied bias voltage should follow 
that of the electrochemical potential of the substrate contact. However, this 
is only true at positive bias voltages. At negative bias voltage, they 
may deviate from each other as a function of the metal-molecule distance. 
This can only be explained from an atomi-level analysis of the charge and 
potential response of the molecule to the applied bias voltage. 

The spatial distribution of the charge perturbation and the voltage drop 
across the molecular junction for $\Delta L=1.5(\AA)$ at voltages 
of $V=2.0(V)$ and $-2.0(V)$ are shown in Fig. (\ref{xueFig2-4}). 
The spatial distribution of the charge perturbation is obtained 
by integrating the difference in electron density at 
finite and zero biases along the z-axis and plotted as a function of 
position in the xy-plane (defined by the benzene ring). The voltage  
drop is obtained by evaluating the difference between the electrostatic 
potential at finite and zero biases, which obeys the boundary condition 
of approaching $-V/2$ ($V/2$) at the substrate (top) electrode. Due to the 
increased metal-molecule distance at the top contact, a larger potential 
barrier is created at the top metal-molecule interface than that at the 
substrate-molecule interface, and it is easier for the electrons to 
move from the top contact side to the substrate side than the other 
way around (not shown here). This leads to distinct behaviors in the 
molecular response to the applied bias at different bias polarities. 

At $V=2.0(V)$, electrons are injected into the 
molecule from the top contact, which forms the bottleneck for transport. The 
direction of the applied field favors the electron flow within the molecules. 
Most of the charge perturbation therefore occurs at the top metal-molecule 
interface, with a large decrease in the electron density on the molecule side 
of the top metal-molecule interface (Fig. (\ref{xueFig2-4})). Since 
electrons on the substrate side of the junction can move freely to screen 
effectively the applied field, most of the voltage drops across the 
top metal-molecule interface. At $V=-2.0(V)$, electrons are 
injected from the substrate contact across a smaller barrier. The 
electrons cannot move freely to screen the 
applied field compared to the positive bias case. Correspondingly, the 
charge perturbation on the substrate side of the molecular junction is 
significant and significant voltage drop occurs before reaching the top 
metal-molecule interface. As a result, at 
large top metal-molecule distance, the molecular level 
shift follows that of the substrate electrode at positive bias voltage, but 
shows more complicated pattern for negative bias voltage with larger effect 
on the occupied molecular states. The total number of electrons within 
the molecule decreases with increasing positive bias voltage since it is 
easier to extract electrons from the molecule through the substrate 
contact than to inject electron into the molecule through the top contact. 
But the total number of electrons is approximately 
constant at negative bias polarity. This gives rise to the large asymmetry in 
I-V and G-V characteristics with respect to the bias polarity. 

We can also examine more directly the asymmetry in the device 
characteristics from the voltage dependence of the transmission (TE) 
characteristics at $\Delta L=1.5 (\AA)$ (not shown here).
The conductance reaches its peak when one of the metal Fermi-levels 
moves into alignment with the peak positions in the TE characteristics.  
We find that at large top metal-molecule distance, the conductance reaches 
its peak near zeros bias in the negative bias direction as the right 
metal Fermi-level moves down passing the HOMO level. At positive bias 
voltage, the G-V characteristics probes the states in the HOMO-LUMO gap   
resulting in a much reduced conductance.

\section{The effect of the atomic structure of the metallic 
surface\label{S3}}
 
If the transport through the molecule is measured using a scanning tunneling 
microscope (STM) tip, it would not be appropriate to model the top contact 
as a semi-infinite crystal since atomic-scale structures may exist at the 
tip apex. If the current is measured using atomic-size break junctions,  
both contacts may include atomic-scale structures on their surface. In this 
section, we investigate the effect on current transport due to the presence 
of atomic structures on the metallic surface. We consider two simple 
device models: For model 1, we consider 
the top electrode as composed of one apex atom sitting on top of the 
triangular gold pad of the semi-infinite $<$111$>$ crystal; 
for model 2, we consider both electrodes as composed of one apex 
atom sitting on top of the triangular gold pad of the semi-infinite 
$<$111$>$ crystal. 

The equilibrium energy-level line up (as reflected in the TE and PDOS plots) 
for both models are shown in Fig. (\ref{xueFig3-1}). The calculated I-V 
and G-V characteristics are shown in Fig.\ (\ref{xueFig3-2}).       
We find remarkable similarity between adding a apex atom onto the top 
electrode (model 1) and increasing the top metal-molecule distance in 
determining the equilibrium ``band'' lineup as well as the nonlinear 
transport characteristics. Both cases result in the HOMO level being 
closer to alignment with the metal Fermi-level. Both cases also show 
a similar lineup scheme for the LUMO level. 
Note that the device characteristics for model 1 
show similar feature with that of $\Delta L=1.0( \AA)$ in sec. \ref{S2}, 
although the distance between the end sulfur atom and the semi-inifinite 
crystal surface here would correspond to $\Delta L=2.3( \AA)$. 
A simple explanation of the electronic origin of this similarity is 
as follows: the valence orbital of an isolated gold apex atom is of $s$ type 
which has minimal overlap with the sulfur $\pi$ orbitals on the 
molecule. The hybridization between the apex gold atomic orbitals and the 
gold surface states introduces non-$s$ type symmetry, so the coupling 
between the apex gold atom and the sulfur end atom is stronger than that 
would be obtained for an isolated gold-sulfur bond, but is much weaker than 
that of the reference interfacial configuration. Although this has been 
pointed out before~\cite{Lang002,Datta98}, further insight 
into the problem can only be obtained through an atomic-level analysis of 
junction charge and potential response because the 
simple bonding analysis cannot explain the similarities in the nonlinear 
transport characteristics.  

The potential perturbation upon formation of the molecular junction 
(at equilibrium) for model 1 is shown in Fig.\ (\ref{xueFig3-3}), while the 
charge and potential response for bias voltages of $-2.0(V)$ and $2.0(V)$ 
are shown in Fig.\ (\ref{xueFig3-4}), where we have also shown the position 
of the molecule and the gold apex atom.  Compared to the molecule-substrate 
contact, the electron density of the equilibrium junction decreases 
in the $p_{x}$ orbital region of the top apex atom, 
since the gold apex $p_{x}$ orbital cannot form a bond with the end 
sulfur atom but is involved in the bonding between the apex atom and 
the top surface atoms (not shown here). This difference 
in the charge perturbation at the two interfaces leads to asymmetry in 
the potential for the equilibrium junction. Although the 
potential change within the molecule favors electron flow from the 
substrate side to the top contact side, the barrier at the 
apex atom-molecule interface is still larger than that at the 
substrate-molecule interface. This 
leads to similar charge and potential response to the applied bias 
voltages as those described in the previous section where the top 
metal-molecule distance is increased. At positive bias, 
the voltage drop across the molecular junction occurs mostly at the right 
metal-molecule contact, but at negative bias voltage, a significant amount 
of the voltage drop also occurs within the molecule 
(Fig.\ (\ref{xueFig3-4})). Compared to the case of 
increasing top contact-molecule distance, the amount of the voltage drop 
within the molecule core is slightly larger here due to the less 
favorable potential landscape for electron flow within the molecule core. 

The similar charge and potential response also leads to 
similar molecular level shift at different bias polarities 
(not shown here). Note that \emph{voltage also drops from the apex 
atom to the bulk of the electrode} (the voltage should approach $+(-)1.0(V)$ 
as we approach the bulk of the electrode). This is because  
a single gold atom at the apex cannot effectively screen  
the applied field, which behaves more like the other atoms in the molecule. 
This may have important implications in transport measurement using 
STM tips, since sharp atomic-scale structures and correspondingly 
localized electron states can develop at the apex of the metallic tip 
especially for transition metals~\cite{Tsukada}. If a significant 
amount of the voltage drops from the apex to the metal bulk, negative 
differential resistance can be observed for single 
molecules~\cite{Xue991,Lang,Avouris}. 

For junction model 2, the above analysis of charge and potential 
response at the top metal-molecule contact applies also to the 
substrate-molecule contact. Compared to the reference symmetric contact 
configuration, the contact to the electrodes leads only to potential barrier 
at the apex atom-end sulfur interface. The potential landscape within the 
molecule core is rather flat, while for the reference contact configuration 
there is an additional potential barrier 
across the sulfur-benzene bonding region (see {\bf I}). 
As a result, once electrons are injected into the molecule, it would be 
easier for them to be extracted through the other contact. This leads to 
different charge and potential response at finite applied bias: most of the 
charge perturbation and voltage drop occurs at the electron injecting 
side of the molecular junction, yielding results (not shown here) very 
similar to Fig. (\ref{xueFig3-2}). As a result, the direction of the 
molecular level shift with applied bias follows that of the Fermi-level of 
the electron injecting contact at both bias polarities. 

The similarity between junction model 1 and increasing top 
metal-molecule distance by $\Delta L=1.0(\AA)$ is reflected equally in the 
equilibrium transmission characteristics and the corresponding LDOS plot, 
where both the magnitude and the energy-dependence of the transmission 
coefficient as well as the charge distribution of the contact-perturbed 
molecular states show similar behavior. By contrast, in the equilibrium 
transmission characteristics of the junction model 2, the overall 
energy-dependence of the transmission 
coefficient corresponds closely to the reference contact 
configuration being shifted up to move into closer resonance with the metal 
Fermi-level due to the weaker bonding across both metal-molecule 
interfaces. But similar to junction model 1, the double-peak structure 
corresponding to transmission through the LUMO and LUMO+1 has been 
reduced to a single peak. Compared to junction model 1, there is a 
sharp peak in the TE characteristics through the metal-induced-gap states at 
$E=-3.1(eV)$ in the HOMO-LUMO gap. The TE through the 
occupied molecular states are also much larger than that of junction 
model 1, and sharper than the reference configuration. All these 
follow from the symmetry and  
magnitude of the charge distribution of the corresponding molecular 
states (determined from the LDOS). The charge distribution (the LDOS) 
of the LUMO and LUMO+1 levels remain qualitatively the same for the 
reference configuration, increasing top metal-molecule 
distance, junction model 1 and model 2. Insertion of an 
apex atom as well as increasing the metal-molecule distance 
results in much reduced overlap between the LUMO state and the 
metal surface states leading to negligible transmission. The LUMO+1 state 
also shows similar charge distribution in all cases, with large weight on 
both end sulfur atoms. This leads to large TE which 
are not affected as strongly by inserting an apex atom or increasing 
metal-molecule distance as other states. The larger TE  
through the HOMO-LUMO gap and the occupied molecular states for junction 
model 2 is instead due to the symmetry of the junction configuration. For 
junction model 2, the corresponding LDOS shows equally large charge 
distribution at both metal-molecule interfaces, while for junction 
model 1, they show much reduced charge at the 
subtrate-molecule contact. The peaks for model 1 and 2 are sharper 
than the reference configuration because the states are more localized. 
So the transmission through the occupied states is reduced by inserting 
an apex atom at one contact but remains large when apex atoms are inserted 
at both contacts. 

To summarize, the results shown here highlight the important effect that 
atomic-scale electrode structures may have on molecular transport 
characteristics. 

\section{The effect of the molecular adsorption geometry\label{S4}}   

In computational study of molecular junctions, the adsorption 
geometry is often chosen with the sulfur atom in a three-fold site above the 
electrode surface. In this section, we consider a different adsorption 
geometry with the sulfur end atom atop one surface gold atom. Each surface 
gold atom has 
$6$ nearest-neighbors, therefore we include $7$ gold atoms on each metallic 
surface into the ``extended molecule'' (denoted model 3). We also consider 
the effect on current transport when the top metal-molecule distance is 
increased by $1.0( \AA)$ (denoted model 4) for comparison with the results 
in sec. \ref{S2}. 

For device model 3, the coupling between the molecule 
and the electrodes is reduced due to the less favorable orbital overlap 
between the end sulfur atom and the gold surface atom directly underneath 
it. But the reduction is smaller than that in sec.\ \ref{S3} due to the 
molecular orbital overlap with the other $6$ gold atoms on the surface. 
Compared to the case of three-fold adsorption, the magnitude of the charge 
transfer and the potential perturbation are smaller (not shown here). As a 
result, the HOMO level is moved slightly closer to the metal Fermi-level 
for device at equilibrium (Fig. \ref{xueFig4-1}), similar to the effect of 
slightly increasing the metal-molecule distance. This geometry effects 
the molecular states differently depending on their composition. 
For sulfur based states like HOMO, the PDOS is narrowed due to the reduced 
mixing with the metal surface states. For carbon based states like 
LUMO, the PDOS is broadened since the mixing with the metal surface 
states is stronger due to the proximity of more gold surface atoms. The same 
consideration leads to the more broadened PDOS towards the HOMO-LUMO gap for 
HOMO and LUMO. Also notable is that the once pronounced transmission peak 
for tunneling through the metal-induced-gap-states is suppressed since the 
mixing through the end sulfur atom is suppressed. The trend continues 
as the top metal-molecule distance is increased 
(model 4 in Fig. \ref{xueFig4-1}). Again increasing the top metal-molecule 
distance leads to the favorable alignment of the HOMO and increases the 
zero-bias conductance (Fig. \ref{xueFig4-2}).

Due to the broadened density of states in the HOMO-LUMO gap and more 
favorable equilibrium energy level lineup, the low-bias conductance is higher 
and its increase with bias voltage is less steep in the atop adsorption 
geometry (Fig. \ref{xueFig4-2}). The peak positions in the G-V characteristics 
are reached at slightly smaller bias voltage due to the smaller difference 
between the HOMO level and the equilibrium Fermi-level.    
 
\section{The effect of a retained end hydrogen atom\label{S5}}      

For the thiol molecules self-assembled on a gold substrate, it is commonly 
believed that the end hydrogen atom is desorbed during the final stage of the 
self-assembly process~\cite{Xia,Tour98}. But it is not clear whether 
the hydrogen 
atom at the top contact is desorbed after the formation of a stable contact 
since different experimental techniques have been used. In this section 
we investigate whether the presence of an additional end hydrogen atom 
at the top contact affects significantly the transport characteristics. 
The structure of the molecular radical is obtained first by optimizing the 
geometry of the molecule with both end H atoms at the BPW91/$6-31G^{*}$ 
level and then removing the H atom at the substrate side, forming a 
molecular spin doublet (optimizing the geometry of the molecular radical 
directly gives similar results). The hydrogen atom should inhibit the coupling 
between the sulfur atom and the top metal surface. For comparison with 
the results in sec. \ref{S2}, we assume the same 
three-fold sulfur-gold adsorption geometry but with increased sulfur-surface 
distance of $2.2( \AA)$ (the hydrogen-surface distance is $2.0(\AA)$).        

The H-atom addition increases the number of electrons of 
the molecule by one. For this doublet system, 
the three molecular orbitals energetically closest to metal Fermi-level 
(corresponding to the HOMO-1, HOMO and LUMO) for electrons with different spin 
differ, as shown in Fig. (\ref{xueFig5-1}), which are also different 
from those of the molecule with both end H atoms removed. The H addition 
breaks the symmetry of the molecular orbitals. Both the HOMO for spin-up 
electron and the HOMO-1 for spin-down electron have large weight only on 
the left (substrate) sulfur and therefore can couple strongly only to the 
left (substrate) contact. 
On the other hand, both the HOMO-1 for spin-up electron and the HOMO for 
spin-down electron show electrons delocalized throughout the molecule, with 
larger weight on the left sulfur for the spin-down electron. The LUMO for 
both spin-up and spin-down electrons shows large weight only on the 
interior carbon atoms, not affected by the H addition. 

The H addition at the top (right) metal-molecule interface saturates 
the sulfur $\pi$ bond and reduces significantly the amount of charge 
transfer into the top sulfur atom upon adsorption onto the electrodes 
due to the saturated bonding. The decreased molecule-metal bonding leads to 
a larger potential barrier at the top interface. 
Indeed, the magnitude of the charge transfer at 
the top contact is smaller than that obtained in sec.\ \ref{S2} for the case 
of $\Delta L=2 (\AA)$, which corresponds to a sulfur-surface distance 
of $3.9(\AA)$. This leads to quite different energy-level lineup scheme for 
the molecular junction at equilibrium, as shown in Fig. (\ref{xueFig5-2}) 
where we plot TE and the PDOS in the molecule. The peak positions 
corresponding to transmission through the HOMO and LUMO levels are 
lowered relative to the reference structure, giving a more symmetric 
location of the metal Fermi-level, although the HOMO level 
is still closer to the Fermi level than the LUMO. 

Note that although both the energy and the 
electron distribution associated with the molecular states depend on the spin 
direction for the isolated molecule, the transmission coefficient and 
the PDOS in the molecular junction \emph{are identical for both spin 
directions once the self-consistent calculation is converged}. The reason 
is as follows: For the strong molecule-metal coupling regime we consider 
here, the open-shell PDT molecular radical (with one H end) within the 
molecular junction is only part of a large quantum system. Since the filling 
of the electron states is determined by the Fermi-distribution of the 
semi-infinite electrodes which are non-magnetic, there is no physical origin 
for breaking the spin symmetry of the system. The PDT molecular radical in 
contact with two gold electrodes behaves in much the same way as an 
open-shell atom within a closed-shell molecule since the electron states are 
delocalized across the molecular junction. The situation will be dramatically 
different in the Coulomb blockade regime when the molecule is weakly 
coupled to both electrodes or when a high spin-degeneracy atom is 
inserted into the molecule~\cite{Mccuen02,Park02}, in which cases 
a spin-dependent interaction term needs to be included in 
the Hamiltonian describing the molecular junction and spin-dependent electron 
scattering can dominant especially at low temperature.      
  
Interestingly, although the addition of a H atom moves the frontier 
orbitals closer to the metal Fermi-level and to each other 
(Fig.\ (\ref{xueFig5-1})), the transmission 
characteristics of the equilibrium junction remain qualitatively the 
same as for the singlet molecular biradical besides the shift in the 
peak positions(Fig.\ (\ref{xueFig5-2})). The corresponding 
LDOS shows that the charge distributions associated with both the HOMO and 
LUMO levels remain similar in shape with the addition of the end H atom. 
The transmission coefficient at the Fermi level is slightly reduced, but 
the overall transmission characteristics in the HOMO-LUMO gap are similar 
in both cases. The main effect of introducing the end H atom has been in 
creating a nonsymmetric device structure. 

The calculated I-V and G-V characteristics are shown in 
Fig. (\ref{xueFig5-3}). Both the current and the conductance are reduced 
by the presence of the hydrogen within the bias range studied. 
Substantial asymmetry with respect to the bias polarity is introduced due 
to the the different contact configuration. Similar to the results 
discussed in previous sections, peak in the conductance is reached only 
at negative bias polarity due to the larger voltage drop at the top 
metal-molecule contact at positive bias (not shown here). This can be 
seen clearly from the bias-dependence of the transmission characteristics in 
Fig. (\ref{xueFig5-4}). Applying a negative bias decreases the difference 
between the HOMO level position of the molecular radical (one H end) and the 
molecular biradical, leading to the same bias voltage of $V=-1.6(V)$ 
where conductance reaches its peak (the peak position in the transmission 
characteristics coincides with the fermi level of the top contact). The shift 
with applied bias of the LUMO level is much larger than that of the HOMO, 
but since the conductance is again determined mainly by the HOMO states, 
this doesn't affect the calculated transport characteristics.            
  
\section{Conclusions \label{S6}}

We have investigated the effect on molecular transport due to different 
structural aspects of the metal-molecule interface . For a given 
metal-molecule combination, the differences in the interface structure 
not only lead to different metal-molecule coupling, but also to 
different energy-level lineup and to different 
electrostatic potential profile across the the molecular junction. The 
difference in the resulting nonlinear transport characteristics reflects 
both the difference in the energy-level lineup scheme as well as the 
difference in the potential response of the molecular junction.  

These considerations are illustrated through detailed microscopic 
calculation of the prototypical molecular device formed by sandwiching 
the PDT molecule between two gold electrodes. The metal-molecule interface 
structures investigated differ in metal-molecule distance, 
atomic structures at the metal 
surface, adsoprtion geometry and the presence of an additional end hydrogen 
atom. The chosen system is representative of the current experimental work 
on molecule electronics, but is also unique in that the HOMO level of 
the isolated PDT molecule is very close to the gold Fermi-level. This 
leads to a counterintuitive increase of conductance with increasing top 
metal-molecule distance beacuse the reduced coupling leads to closer 
alignment of the HOMO level with the gold Fermi-level. The conductance 
decreases only after passing the maximum metal-molecule distance where 
the energy level lineup becomes essentially identical to that of 
the molecule chemisorbed on the substrate. For the gold-PDT-gold junction 
we consider here this happens around $\Delta L=1.5 (\AA)$ corresponding to 
the sulfur-top metal distance of $3.4 (\AA)$. 
Adding one apex atom onto the semi-infinite surface of the bulk 
electrodes is equivalent in its effects to the increase of the 
metal-molecule distance due to the unfavorable orbital overlap with 
the sulfur end atom. The similarity is reflected in the contact-perturbed 
molecular states, the charge and potential response of 
the molecular junction to the applied bias and the 
nonlinear transport characteristics. Changing to an atop molecular adsorption 
geometry leads to slightly favorable energy level lineup 
for the molecular junction at equilibrium and consequently larger 
conductance, but the overall transport characteristics remain qualitatively 
the same. The presence of the additional hydrogen end atom at the top 
metal-molecule contact affects substantially the electronic 
processes in the molecular junction due to the different nature of the 
molecular orbitals and the asymmetric device structure involved, 
reducing the conductance and current compared to the case where the 
end hydrogen atom is desorbed during the formation of the contact. 
The results of the microscopic calculation can all be understood 
from the equilibrium energy-level lineup combined with qualitative 
knowledge of the voltage drop due to the asymmetry in the two 
metal-molecule contacts. Since the equilibrium energy level lineup is 
quite sensitive to the atomic-scale structures of the metal-molecule 
interface, a correct identification of the device structure is essential 
for explaining the transport characteristics of any molecular junction. 
  
For different metal-molecule combinations, the lineup of the molecular level 
with respect to the metal surface band, and therefore the functional 
dependence of transport characteristics on interface structure 
will be different. In particular, both the energy level lineup and 
the metal-molcule coupling can be modified by simply replacing the end 
contact atom (sulfur in this paper) with other chemical groups without 
changing the metal and the molecule core. This and other chemical aspects 
of the metal-molecule interface are interesting subjects which deserve 
further analysis.  

\begin{acknowledgments}
We thanks A. Nitzan, S. Datta, V. Mujica and H. Basch for useful discussions.
This work was supported by the DARPA Molectronics program, the DoD-DURINT 
program and the NSF Nanotechnology Initiative. 
\end{acknowledgments}





%


\newpage
  
\begin{figure}
\caption{\label{xueFig2-1-1} 
Transmission versus energy (TE) and projected density of states (PDOS) in 
units of (1/eV) corresponding to the HOMO and LUMO as a function of the 
top metal-molecule distance $\Delta L$ at 
$\Delta L=0.5,1.0( \AA)$. For comparison, 
we have also shown the TE and PDOS characteristics for the reference device 
structure where the molecule forms symmetric contact with the two electrodes 
(solid line). The horizontal line in the TE plot shows the Fermi-level 
position. The horizontal lines in the PDOS plot show the energies of the 
HOMO and LUMO levels in the isolated molecule. }
\end{figure}   

\begin{figure}
\caption{\label{xueFig2-1-2} 
Transmission versus energy (TE) and projected density of states (PDOS) in 
units of (1/eV) corresponding to the HOMO and LUMO as a function of the 
top metal-molecule distance $\Delta L$ 
at $\Delta L=1.5,2.0( \AA)$. For comparison, 
we have also shown the TE and PDOS characteristics for the reference device 
structure where the molecule forms symmetric contact with the two electrodes 
(solid line). The horizontal line in the TE plot shows the Fermi-level 
position. The horizontal lines in the PDOS plot show the energies of the 
HOMO and LUMO levels in the isolated molecule. }
\end{figure}   

\begin{figure}
\caption{\label{xueFig2-2} 
Current-voltage (I-V) and conductance-voltage (G-V) characteristics of the 
gold-PDT-gold junction as a function of the top metal-molecule distance. 
For comparison, we have also shown the I-V and G-V characteristics for 
the reference device structure where the molecule forms symmetric contact 
with the two electrodes (solid line). }
\end{figure}   

\begin{figure}
\caption{\label{xueFig2-3}
Bias-induced modification of molecular levels as a function of top 
metal-molecule distance. We have also shown the 
position of the equilibrium Fermi-level $E_{F}$ and the electrochemical 
potential of the two electrodes $\mu _{L(R)}$ in the plot.}
\end{figure}   

\begin{figure}
\caption{\label{xueFig2-4}
Spatial distribution of charge transfer and potential drop at bias voltage 
of 2.0(V) at the gold-PDT-gold contact for $\Delta =1.5(\AA)$. 
The upper figure shows the 
difference between the electron density at finite bias 
voltage and the electron density for device at equilibrium, the lower 
figure shows the difference between the electrostatic potential at that 
voltage and the electrostatic potential for device at equilibrium.}
\end{figure}   

\begin{figure}
\caption{\label{xueFig2-5} 
Three dimensional plot of the bias-dependence of the transmission versus 
energy characteristics for $\Delta =1.5(\AA)$. The two lines in the X-Y 
plane show the electrochemical potential of the two electrode 
$\mu _{L(R)}$ as a function of applied bias voltage.}
\end{figure}  
   
\begin{figure}
\caption{\label{xueFig3-1} 
Transmission versus energy (TE) and projected density of states (PDOS) in 
units of (1/eV) 
corresponding to the HOMO-1, HOMO, LUMO and LUMO+1 for device model 1 and 
model 2. The horizontal line in the TE plot shows the Fermi-level 
position. The horizontal lines in the PDOS plot show the energies of the 
HOMO-1, HOMO, LUMO and LUMO+1 levels in the isolated molecule.}
\end{figure}   

\begin{figure}
\caption{\label{xueFig3-2} 
Current-voltage (I-V) and conductance-voltage (G-V) characteristics of the 
gold-PDT-gold for device model 1 and 2. }
\end{figure}   

\begin{figure}
\caption{\label{xueFig3-3}
Electrostatic potential change upon the formation of the 
molecular junction for model 1 as a function of position in the xy-plane. 
Also shown is the projection of the molecule and the apex atom 
onto the xy-plane.}
\end{figure}   

\begin{figure}
\caption{\label{xueFig3-4}
Spatial distribution of charge transfer and potential drop at bias voltage 
of $-2.0(V)$ and $2.0(V)$ for device model 1. }
\end{figure}   

\begin{figure}
\caption{\label{xueFig4-1} 
Transmission versus energy (TE) and projected density of states (PDOS) in 
units of (1/eV) 
corresponding to the HOMO and LUMO for the atop adsorption geometry 
with symmetric contact (model 3) and with increased top metal-molecule 
distance (model 4, $\Delta L=1.0(\AA)$). }
\end{figure}

\begin{figure}
\caption{\label{xueFig4-2} 
Current-voltage (I-V) and conductance-voltage (G-V) characteristics of the 
gold-PDT-gold junction as a function of molecular adsorption geometry. }
\end{figure}   

\begin{figure}
\caption{\label{xueFig5-1}
Orbital shape of the HOMO-1, HOMO and LUMO states of PDT molecule with an 
additional hydrogen end atom for both spin-up and spin-down electrons.}
\end{figure}   

\begin{figure}
\caption{\label{xueFig5-2} 
Transmission versus energy (TE) and projected density of states (PDOS) in 
units of (1/eV) 
corresponding to the HOMO-1, HOMO, LUMO and LUMO+1 for the molecule 
with the end hydrogen atom.  The horizontal line in the TE plot shows the 
Fermi-level position. The horizontal lines in the PDOS plot show the 
energies of the HOMO and LUMO levels for spin-up electrons in the 
isolated molecule. }
\end{figure}

\begin{figure}
\caption{\label{xueFig5-3} 
Current-voltage (I-V) and conductance-voltage (G-V) characteristics 
for the molecule with the end hydrogen atom. }
\end{figure}   
  
\begin{figure}
\caption{\label{xueFig5-4} 
Three dimensional plot of the bias-dependence of the transmission versus 
energy characteristics for the molecule with the end hydrogen atom. 
The two lines in the X-Y plane show the electrochemical 
potential of the two electrode $\mu _{L(R)}$ as a 
function of applied bias voltage.}
\end{figure}  
 
\end{document}